\documentclass[graybox]{svmult}

\usepackage{mathptmx}       
\usepackage{helvet}         
\usepackage{courier}        
\usepackage{type1cm}        
\usepackage{makeidx}         
\usepackage{graphicx}        
\usepackage{multicol}        
\usepackage[bottom]{footmisc}

\makeindex             

\begin{document}

\title*{Local moment approach to multi-orbital Anderson and Hubbard models}
\author{Anna Kauch and Krzysztof Byczuk}
\institute{Anna Kauch \at Institute of Theoretical Physics,
University of Warsaw , ul. Ho\.za 69, PL-00-681 Warszawa, Poland, \email{anna.kauch@fuw.edu.pl}
\and Krzysztof Byczuk \at Theoretical Physics III, Center for Electronic Correlations and
Magnetism, Institute for Physics, University of Augsburg, D-86135
Augsburg, Germany, \email{kbyczuk@physik.uni-augsburg.de}}
%
%
\maketitle

\abstract*{The variational local moment approach (V-LMA), being a modification of
the method due to Logan {\it et al}., is presented here.
The existence of  local moments is taken
from the outset and their values are determined through variational
principle by minimizing the corresponding 
ground state energy.
Our variational procedure allows us to treat both fermi- and
non-fermi liquid systems with many orbitals as well as insulators without any additional
assumptions. 
It is proved by an  explicit construction of the corresponding Ward
functional that 
the V-LMA belongs to the class of conserving approximations. 
As an illustration, the V-LMA is used  to solve the multi-orbital single
impurity Anderson model. 
The method is also applied to solve the
dynamical mean-field  equations for the multi-orbital Hubbard
model. 
In particular, the Mott-Hubbard  metal--insulator transition is
addressed within this approach. 
} 

\abstract{The variational local moment approach (V-LMA), being a modification of
the method due to Logan {\it et al}., is presented here.
The existence of  local moments is taken
from the outset and their values are determined through variational
principle by minimizing the corresponding 
ground state energy.
Our variational procedure allows us to treat both fermi- and
non-fermi liquid systems as well as insulators without any additional
assumptions. 
It is proved by an  explicit construction of the corresponding Ward
functional that 
the V-LMA belongs to the class of conserving approximations. 
As an illustration, the V-LMA is used  to solve the multi-orbital single
impurity Anderson model. 
The method is also applied to solve the
dynamical mean-field  equations for the multi-orbital Hubbard
model. 
In particular, the Mott-Hubbard  metal--insulator transition is
addressed within this approach. }

\section{Introduction}
\label{sec:1}
The single impurity Anderson model (SIAM) is one of the most
investigated models in condensed matter physics \cite{anderson}.
This model is regarded as a prototype to understand and describe:
i) properties of metals with  magnetic atoms \cite{hewson}, ii)
charge transport through quantum dots \cite{dots}, iii) Mott-Hubbard
metal-insulator transitions 
(MIT) within the dynamical mean-field theory (DMFT)
\cite{vollhardt,bulla_2006,georges,pruschke95,vollhardt93,metzner89},
and iv) a crossover  
between weak and strong coupling limits and confinement phenomena. 
The SIAM consists of a term describing band electrons coupled by 
hybridization to a term corresponding to a single impurity where the
local Coulomb interaction is taken into account \cite{anderson}.  
In the featureless hybridization  limit the SIAM is solved exactly
within the Bethe ansatz or conformal field theory techniques so the
ground state and the whole excitation spectrum as well as
thermodynamics are exactly known \cite{hewson}.  
Unfortunately, these methods cannot in practice provide 
dynamical quantities, for example one-particle spectral
functions or dynamical susceptibilities, for all interesting energies. 
Also the (asymptotic) exact solvability is not possible for a general
hybridization term. 

For practical applications of the SIAM one has to rely on either a
numerically exact or an analytical but approximate solution. 
Numerically exact methods, like the numerical renormalization group (NRG) 
\cite{bulla} or the determinant quantum Monte Carlo (QMC) \cite{georges}
are very time (CPU) consuming. 
In particular, the CPU is very long when the
number of orbitals is large in the NRG case and when the temperature
is low in the QMC case. 
Also to extract dynamical quantities is a rather tricky task \cite{max}. 
Reliable analytical methods are therefore needed. 
One of such methods, which recovers properly both weak and strong
coupling limits, is  a {\it local moment approach} (LMA)
 invented recently by Logan {\it et  al.} \cite{Logan}. 

The LMA is a perturbative method around an unrestricted Hartree-Fock
solution with broken symmetry, i.e. with a non-zero local magnetic
moment. 
The broken symmetry is restored at the end by taking  the
average of the solutions corresponding to different  directions of the
local magnetic moment \cite{Logan}.  

In the present contribution we describe the LMA method and our
implementation of it, which is different from the original one \cite{Logan}
by the way of how the value of the local moment is determined. 
Namely, we use the variational principle demanding that the ground
state energy is minimized by the physical value of the local
moment. 
Therefore we use the name variational local moment approach (V-LMA)
for this method. 
Such a procedure allows us to easily generalize the V-LMA for multi-orbital
models as well as for finite temperatures and systems with disorder
\cite{byczuk1,byczuk2,byczuk3,byczuk4}.  
We also discuss the Luttinger-Ward generating functional for the V-LMA
and claim that this method belongs to the class of conserving
approximations. 
The application of LMA for studying  the
electron flow through quantum dots and the Mott-Hubbard MIT is
addressed at the end of the contribution.

\section{Local moment method in one orbital SIAM}

The single impurity Anderson model is given  by the Hamiltonian 
\begin{equation}
H_{\rm{SIAM}}=H_{c}+H_{imp}+ H_{hyb},
\end{equation}
where the conduction electrons are described by 
\begin{equation}
H_{c}=\sum_{\bf{k},\sigma} \epsilon_{\bf k} c_{{\bf k}\sigma}^{\dagger} c^{\phantom\dagger}_{{\bf k}\sigma},
\end{equation}
where $\epsilon_{\bf k}$ is an energy (a dispersion relation) for an
electron in a state $\bf k$ and spin $\sigma=\pm1/2$, 
the impurity electrons with the local Coulomb interaction $U$ are represented by  
\begin{equation}
H_{imp}=\sum_{\sigma}\left(\epsilon_{d}+U
n_{d-\sigma}\right) n_{d\sigma},
\end{equation}
with $n_{d\sigma}= d_{\sigma}^{\dagger}d_{\sigma}$,
and the hybridization between conduction and impurity electrons is
\begin{equation}
H_{hyb}=\sum_{{\bf k},\sigma}\left( V_{{\bf k}}
d_{\sigma}^{\dagger} c^{\phantom\dagger}_{{\bf k}\sigma} + h.c.\right).
\end{equation}
All local (on impurity site) properties are expressed by the
hybridization function 
\begin{equation}
\Delta(\omega)=\sum_{\bf k}\frac{|V_{\bf k}|^2}{\omega-\epsilon_{\bf k}},
\end{equation}
and not by $\epsilon_{\bf k }$ and $V_{\bf k}$ separately. 
This can be proved by tracing out the non-interacting conducting
electrons. 

\subsection{Mean field solution of the single impurity Anderson  model} 

The Hartree-Fock mean-field solution of the SIAM is obtained by
factorizing the interacting term 
$n_{d\uparrow}n_{d\downarrow}\approx
\langle n_{d\uparrow}\rangle n_{d\downarrow}+n_{d\uparrow} \langle
n_{d\downarrow}\rangle - \langle n_{d\uparrow}\rangle \langle
n_{d\downarrow}\rangle$ 
\cite{anderson}. 
For the interaction $U$ above  $U_c$ and corresponding impurity
electron densities $\bar{n}_{d}$ the mean-field solution is unstable
toward the local moment formation with non-zero  moment $\mu
\equiv \langle n_{d\uparrow}\rangle - \langle n_{d\downarrow}\rangle $.  
The solution is doubly degenerate because of two equivalent
directions of the local moment $\mu = \pm |\mu|$, which give the same
energy of the system. 
The local (impurity) Green function within  the Hartree-Fock solution is
\begin{equation}
G_{\sigma}^{HF}(\omega)=  \frac{1}{\omega -\epsilon_d-
  \Delta(\omega)-\Sigma^{HF}_{\sigma}+ i\delta {\rm sgn}\omega}  
\end{equation}
where the static Hartree-Fock self-energy
$\Sigma_{\sigma}^{HF}=U\langle n_{\bar{\sigma}}\rangle$ and $\delta\rightarrow 0^+$.  
Since there are in principle two possible signs
of the local moment, there are two different possible Hartree -- Fock
Green functions denoted by $G^A_{\sigma}(\omega)^{HF}$ and
$G^B_{\sigma}(\omega)^{HF}$ that differ only by the sign of the local
moment and depend parametrically on its value  $|\mu|$.

The fundamental deficiency of the Hartree-Fock approximation is
that it leads to a broken symmetry solution which cannot persist in
the thermodynamic limit, i.e. a single impurity cannot lead to the
magnetic solution in the infinite system. 
Also this solution does not  recover the singlet ground state known
from the exact Bethe ansatz solution. 
Nevertheless it turns out to be useful as a starting point in the further
perturbative calculation combined with the  symmetry restoration.

\subsection {Two self-energy description}

\begin{figure}
     \centering
     \includegraphics[width=0.85\textwidth]{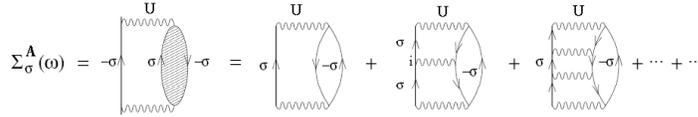}
     \caption{The frequency dependent part of self-energy expressed as
       the RPA series around the broken symmetry Hartree-Fock
       solution. The transverse spin polarization bubbles constitute a
       geometric series which can be summed up to infinity.  } 

\label{fig1}
 \end{figure}

The two Hartree-Fock Green functions 
$G^{A,B}_{\sigma}(\omega)^{HF}$
are used in the time-dependent many-body perturbation expansion. 
Within  the random phase approximation (RPA) the
polarization diagrams are

\begin{equation}
\Pi^{AA}_{\sigma\bar{\sigma}}( \omega)=\frac{^0\Pi^{AA}_{\sigma\bar{\sigma}}( \omega)}{1-U^0\Pi^{AA}_{\sigma\bar{\sigma}}( \omega)}
\end{equation}
and correspond to spin flip processes as represented by the
 Feynmann diagrams in Fig.~\ref{fig1}.

For each type of the mean-field solution $G^{A,B}_{\sigma}(\omega)^{HF}$ we
have the corresponding self-energy 
\begin{equation}
\Sigma^{A}_{\sigma}(\omega)= \Sigma_{\sigma}^{HF}+U^2 \int \frac{d\omega'}{2\pi i} \Pi^{AA}_{\bar{\sigma}\sigma}( \omega')G^{A}_{\bar{\sigma}}(\omega-\omega')^{HF} 
\end{equation}
depending on frequency and  parametrically on $|\mu|$ as well.
The full RPA-Green functions $G^{A,B}_{\sigma}(\omega)$ are constructed by 
using the Dyson equation separately for A and B solutions. 
Note that $G^{A,B}_{\sigma}(\omega)$ depends parametrically on still
unknown $|\mu|$.

\subsection {Symmetry restoration ansatz}

To restore the spin-rotational  symmetry Logan {\it et al.}
\cite{Logan} proposed the following ansatz for the full  symmetrized  
Green function 
\begin{equation}
G_{\sigma}(\omega)= \frac{1}{2} \left ( G^{A}_{\sigma}(\omega)+
  G^{B}_{\sigma}(\omega)\right ). 
\end{equation}
Within the LMA the physical Green function is an average of the two
solutions with equal probabilities. 
Although each $G^{A,B}_{\sigma}(\omega)$ is determined within the
renormalized perturbation scheme the final Green function turns out to
capture nontrivial non-perturbative physics as was shown by Logan {\it
  et al.} \cite{Logan} and is also reproduced  below. 
In particular, the LMA is able to recover the Kondo peak in the
spectral function correctly with the exponential width. 

\subsection {Determining the value of local moment}

The value of the local moment is a free parameter and  must still be 
determined. 
In the original approach, Logan {\it et al.} \cite{Logan} imposed the
Fermi liquid 
condition to determine $|\mu|$ at zero temperature. 
This condition might be too restrictive at finite temperatures or in
the multi-orbital cases. 
Therefore we decided to find the physical solution to the problem by  
minimizing the relevant thermodynamical potential with respect to
$|\mu|$ \cite{kauch}. 
At zero temperature the relevant potential is just the ground state
energy of the system, i.e. 
\begin{equation}
E_{\rm{physical}} = \min_{\left\{\mu,n\right\}} E_G(\mu,n),
\end{equation}
where in the case away of half-filling the particle density $n$ must
also be determined.  
The variational method reproduces the Fermi liquid
properties where they are expected.

\subsection {Ground state energy in the Anderson impurity model} 

The ground state energy of the SIAM is given by $E_G=\langle
0|H|0\rangle $. 
This quantum-mechanical  average consists of two parts: the bulk,
which is proportional to the system volume and is independent of the local
moments, and the impurity part, which depends explicitly on $|\mu|$.
The impurity part of the ground state energy, expressed by the
local Green function $G_{\sigma}(\omega)$ and the hybridization
function $\Delta(\omega)$, is equal to \cite{ee}
\begin{eqnarray}
E_{imp}= \frac{1} {2\pi i} \sum_{\sigma} \oint_C d \omega \left[
  \frac {\omega + \epsilon _d + \Delta (\omega)}{2} 
 -\omega \frac{\partial \Delta (\omega)}{\partial \omega} \right]
G_{imp}^{\sigma} (\omega), 
\end{eqnarray}
where the contour integral is over the half circle in the upper
complex plane.  

\subsection{LMA as a conserving approximation}

According to Kadanoff and Baym \cite{Kadanoff} any approximate theory is
conserving if there exists a Luttinger-Ward functional ${\rm \Phi}[G]$
for this theory. 
It is necessary that this functional: i) is universal, i.e. it
dependents only on the full propagator $G_{\sigma}(\omega)$ and not on
the atomic properties of the system and ii) has a functional derivative with
respect to $G_{\sigma}(\omega)$ which is by definition equal to the
self-energy of the system. 
It can be shown \cite{kauch} that the LMA is a conserving approximation and we can
construct explicitly the Luttinger-Ward functional 
\begin{eqnarray}
{\rm \Phi}[G] = 
{\rm \Phi}[G^A,G^B]=\frac{1}{2}\left({\rm \Phi}_{RPA}^A + {\rm
    \Phi}_{RPA}^B \right )
+ \frac{1}{2} Tr \log G_{\sigma}^A G_{\sigma}^B   + \nonumber \\
 - Tr \log \left (
  \frac{1}{2} \left (G_{\sigma}^A + G_{\sigma}^B \right ) \right),  
\end{eqnarray}
where $Tr=T\sum_{\sigma} \sum_{i\omega_n} e^{i0^+}$ and the
functionals  ${\rm \Phi}_{RPA}^{A,B} $ are represented dia\-gram\-mati\-cally
by the RPA diagrams with $G_{\sigma}^{A,B}(\omega)$ respectively.
The constraint that  $G=\frac{1}{2}\left( G_A+G_B \right )$ must be satisfied.
Finally, the free energy functional is given by 
\begin{equation}
\Omega[G] = \Phi[G] + Tr \log G - Tr \Sigma G
\end{equation}
and the stationarity condition $\delta \Omega [G]/\delta G$ gives the
Dyson equation and the physical solution for $G$.

The fact that the LMA is a conserving approximation, as we proved
above,  makes this theory
reliable in describing correlated electron systems, in particular in
the intermediate regimes of parameters.

\section {Local moment approach  for  the multi-orbital SIAM}

In reality the magnetic impurities in metals are atoms with partially
field d- or f-orbitals. 
Such orbitals have degenerate levels. 
Even when a particular environment which
decrease the symmetry and leads to $e_g$ and $t_{2g}$ split levels,
partial degeneracy between orbitals remains. 
The appropriate model to describe such situations is the multi-orbital
single impurity Anderson model. 
It describes a single impurity with many orbital levels $\alpha$, which can be
degenerate or split depending on the single-body matrix element
$\epsilon_{\alpha}$. 
In this case the electrons can interact via direct (density-density)
type of the interaction and via the exchange (Hund) interaction. 
Microscopically, the single impurity Anderson model with many orbital
levels is given by the Hamiltonian: 
\begin{eqnarray}
H_{\rm{SIAM}}=\sum_{\alpha,\sigma}\left(\epsilon_{\alpha}+U_{\alpha}
n_{\alpha,\bar{\sigma}}\right) n_{\alpha,\sigma}+ 
\sum_{\sigma,\sigma'}\sum_{\alpha\neq\beta}\left( U'_{\alpha\beta}
-J\delta_{\sigma\sigma'}\right)
n_{\alpha\sigma}n_{\beta\sigma'}+\nonumber  \\
+ \sum_{{\bf k},\sigma,\alpha} V_{{\bf k}\alpha}\left(
d_{\alpha\sigma}^{\dagger} c_{{\bf k}\sigma}^{\phantom\dagger} + c_{{\bf
    k}\sigma}^{\dagger}d_{\alpha\sigma}^{\phantom\dagger} 
\right) +
\sum_{\bf{k},\sigma} \epsilon_{\bf k} c_{{\bf k}\sigma}^{\dagger}
c_{{\bf k}\sigma}^{\phantom\dagger},
\end{eqnarray}
where the direct $U$ and $U'$ as well as exchange $J$ interactions
between the electrons of spin $\sigma$ and on orbitals $\alpha$ or
$\beta$ are taken into account.  

This multi-orbital version of the SIAM is also of interest in quantum
dot physics, where dots with a few orbitals can be prepared and
investigated experimentally. 
One of the interesting aspect of such system is the possibility to
observe the orbital Kondo effect \cite{orbital_kondo}. 

\subsection {LMA generalization}
In the mean field approximation of the multi-orbital SIAM we also
encounter a doubly degenerate solution, where the two possible Green
functions differ only by the sign of the impurity magnetic moment. 
Within the LMA, we introduce for each pair of orbital indices
$\alpha$ and  $\beta$ the two Green functions $G^{\alpha
  \beta,A}_{\sigma}(\omega)^{HF}$ and $G^{\alpha
  \beta,B}_{\sigma}(\omega)^{HF}$  that correspond to the two 
possible directions  of the total magnetic moment on the impurity. 
These Hartree-Fock Green functions depend now parametrically on values
of local moments on each of the orbitals $\mu_{ \alpha}$. 
Next we use the RPA approximation  to obtain two 
Green functions   
$G^{\alpha \beta,A}_{\sigma}(\omega)$ 
and 
$G^{\alpha  \beta,B}_{\sigma}(\omega)$, 
which are parametrically dependent on the local moments on each orbitals
$\mu_{\alpha}$.   

\subsection {Symmetry restoration and determining the local moment values}

The symmetry restoration in the multi-orbital case is a
straightforward generalization of the previous ansatz, i.e.  
\begin{equation}
G^{\alpha\beta}_{\sigma}(\omega)=\frac {1}{2} \left (G^{\alpha\beta,
    A}_{\sigma}(\omega)+G^{\alpha\beta, B}_{\sigma}(\omega) \right ), 
\end{equation}
except that now the  symmetrized Green functions
$G^{\alpha\beta}_{\sigma}(\omega)$ depend explicitly on local moments
on all of the orbitals, i.e. $|\mu_{ \alpha}|$. 
The parameters $|\mu_{ \alpha}|$ have to be determined independently.
They are  found by the minimization of the ground state energy of 
the impurity with respect to both 
local moment values on orbitals $\mu_{ \alpha}$ and particle number on
each of the orbitals  $n _{\alpha}$ 
\begin{equation}
E_{\rm{physical}} = \min_{\left\{\mu_{\alpha},n_{\alpha}\right\}}
E_G(\mu_{\alpha},n_{\alpha}). 
\end{equation}
As mentioned above, the variational procedure allows us to extend the
LMA on the multi-orbital cases, where the Luttinger (Fermi liquid)
condition for each orbital is absent. 
Also the possibility of non-Fermi liquid solution is naturally
included within present  generalization of the LMA \cite{kauch}, 
i.e. the variational local moment approach.

\section{Application to multilevel quantum dots}

A single  quantum dot  with many atomic-like levels coupled to leads
are described by a multi-orbital single impurity Anderson model:
$$
H=H_{\rm{dot}}+H_{\rm{leads}}+H_{\rm{dot-leads}},
$$
where $H_{\rm{dot}}$ is the local impurity part of the SIAM
Hamiltonian, $H_{\rm{leads}}$ corresponds to the conduction electron
part of the SIAM Hamiltonian, and   $H_{\rm{dot-leads}}$ is equal to
the hybridization term in SIAM \cite{dots}. 

\subsection{V-LMA in quantum dots}

\begin{figure}
     \centering
     \includegraphics[width=0.5\textwidth]{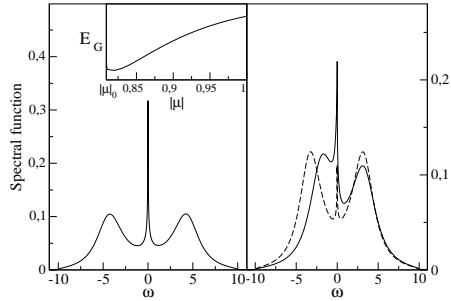}

 \caption{Spectral functions for one level (left panel) and two level
   (right panel) quantum dots.  
Left panel: spectral function at half-filling and $U=6$ (inset: the
ground state energy as  a function of  the absolute value of local
moment $|\mu|$; the axis starts at the Hartree-Fock value
$|\mu|=|\mu_0|$). Right panel:  
   orbitally resolved spectral functions in the dot for  $U=3$,
   $J=0.25U$,  $|\epsilon_{1}-\epsilon_{2}|=0.1U$, and the total
   filling $n_d=1.95$. All curves are for semi-elliptic hybridization
   function with the width $W=20$. The Fermi level is at zero.}
\label{fig2}
 \end{figure}

The properties of transport in a quantum dot in equilibrium, i.e. with
infinitesimally small bias voltage between the leads, 
are determined by  the spectral functions on each of
the orbitals. 
Examples of the spectral functions are presented in
Fig.~\ref{fig2} for the one-orbital case (left panel) and for the two
orbital case (right panel). 
In the two orbital case the atomic levels are shifted such that one of
the orbitals is at half filling (dashed line) and the other is away of
half filling (solid line). 
The Kondo peak in the symmetric case is suppressed by the exchange
(Hund) interaction ($J\ne 0$), which favors parallel spin orientations. 
In the asymmetric case the Kondo peak survives due to the presence of
uncompensated magnetic moment and is shifted toward  the
lower Hubbard band. 
Further investigation of multilevel quantum dots including transport
properties will be presented elsewhere \cite{kauch}.

\section{Application to the multi-orbital Hubbard model}


The generalized variational LMA is also applied to solve the multi-orbital Hubbard
model 
$$H_{\rm Hubb}=\sum_{ij}\sum_{\alpha,\sigma}t_{ij}^{\alpha}
d^{\dagger}_{i\alpha\sigma} d_{j\alpha\sigma}+H_{\rm local},$$
 where the local part is a lattice sum of the terms which are of the
 same form as the atomic part in  the SIAM.  
This model is solved within the DMFT where the self-consistency condition
relates the local matrix Green functions with the matrix of the
self--energies \cite{georges}. In this way the lattice problem is mapped
onto the Anderson impurity problem which has to be solved for
different hybridization functions until self-consistency is
achieved. In order to solve the Hubbard model within DMFT we need to
solve the SIAM for arbitrary hybridization functions. The 
self--consistency condition simplifies greatly for the Bethe lattice
which is used in this contribution.  

\subsection {V-LMA method in DMFT}

In the recent few years the orbital-selective Mott-Hubbard
metal-insulator transition has been the subject of extensive
studies \cite{koga,liebsch,vandongen,biermann,held}. 

\begin{figure}[tbp]
     \centering
     \includegraphics[width=0.5\textwidth]{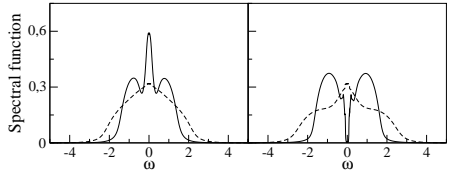}
      \caption{Spectral functions for two-orbital Hubbard model with
        different band widths $W=2$ (solid line) and $W=4$ (dashed
        line) on Bethe lattice with infinite coordination number. Left
        panel:  $U=1.2$, $J=0.1U$. Right panel: $U=2$,
        $J=0.1U$. The inter-band interaction $U'=U-2J$ is fixed
        preserving $SU(4)$ symmetry. In both cases the Fermi level is
        at zero energy and the bands are half filled.} 

\label{fig3}
 \end{figure}

Using the V-LMA to obtain the solution of the SIAM in each of the DMFT loops
the spectral functions for two--orbital Hubbard model at zero
temperature were found. 
As an example, Fig.~\ref{fig3} shows the results for
the case with different bandwidths and non-zero Hund coupling
$J$.
Since one of the spectral function is metallic-like (finite at
$\omega=0$) and the other is insulating-like (vanishes at $\omega=0$)
we conclude that  the orbital selective MIT occurs in this model
system.


At the end we discuss the V-LMA in perspective to other methods used to
solve the impurity problem and DMFT equations. 
The V-LMA belongs to the class of approximate, analytical methods like
for example the iteration perturbation theory (IPT) \cite{ipt}, the non-crossing
approximation (NCA) \cite{nca}, or slave-boson theory (SB) \cite{sb}, and various
extenstions of these methods. 
As we showed here, the V-LMA is a conserving approximation, contrary for
example to the IPT, and correctly describes high- and low-energy parts
of the spectra, recovering the Kondo peak and Luttinger pinning. We
tested this theory at zero temperature but there is no conceptual
obstacle why the V-LMA should not work at finite temperatures as well. 

The V-LMA is not numerically exact like the quantum Monte
Carlo method \cite{georges}, the numerical renormalization group (NRG) \cite{bulla}, dynamical
matrix renormalization group (DMRG) \cite{dmrg}, or exact diagonalization
(ED) \cite{ed}. However, each of the numerically exact methods suffers from
principal obstacles in practical applications, in particular when the
temperature is too low (QMC) or too high (NRG), or number of orbitals
increases (NRG, DMRG, ED). 

Therefore we conclude that the V-LMA is a method of choice for solving
the DMFT equations and can be used as a relatively fast and accurate
impurity solver. 
The only technical difficulty in the variational LMA is to compute
with high accuracy the system energy and to find its minimum. This
should be performed with a great care.

 \section*{Summary}

The generalized variational LMA to the multi-orbital SIAM allows us
to efficiently solve the problems of correlated electron systems such as
multilevel quantum dots and the Hubbard model within the DMFT. In particular 
it is relatively easy to address the problems of different band widths
and also  the removing of the orbital de\-ge\-ne\-ra\-cy \cite{kauch2}.
We  experienced that the local moment approach is an efficient
method in studying these problems, in particular, when the number of the orbitals is
larger than two. \\

\vspace{0.5cm}

\begin{acknowledgement}
We thank dr.~R.~Bulla and  prof.~D.~Vollhardt for the discussions. The
hospitality at Augsburg University is also acknowledged.
This work is supported by Sonderforschungsbereich 484 of the Deutsche
Forschungsgemeinschaft (DFG) and by the European Commission under the
contract No MRTN-CT-2003-504574.\\
\end{acknowledgement}
%

%
%
%

\end{document}